# Max Edge Coloring of Trees

Giorgio Lucarelli*    Ioannis Milis*    Vangelis Th. Paschos†

June 29, 2018

**Abstract**

We study the weighted generalization of the edge coloring problem where the weight of each color class (matching) equals to the weight of its heaviest edge and the goal is to minimize the sum of the colors' weights. We present a 3/2-approximation algorithm for trees.

## 1 Introduction

In the standard *edge coloring* problem we ask for a partition $S = \{M_1, M_2, \ldots, M_s\}$ of the edge set of $G$ into color classes (matchings) such that $s$ is minimized. In this parer we study the following generalization of the standard edge coloring problem which arises in the domain of optical communication systems (see for example [4]): A positive integer weight is associated with each edge of $G$ and we now ask for a partition $S = \{M_1, M_2, \ldots, M_s\}$ of the edges of $G$ into color classes, each one of weight $w_i = \max\{w(e)|e \in M_i\}$, such that the sum of the colors' weights $W = \sum_{i=1}^{s} w_i$ is minimized.

The analogous generalization for the standard vertex coloring problem, where weights are associated to the vertices of a graph and the weight of each color class (independent set) equals to the weight of its heaviest vertex, has been also addressed in the literature and it is known as Max (Vertex) Coloring (MVC) problem [8, 7]. Respectively to this we refer to our problem as Max Edge Coloring (MEC) problem.

It is known that the MEC problem is strongly NP-hard and 7/6 inapproximable even for cubic planar bipartite graphs with edge weights $w(e) \in \{1, 2, 3\}$ [1]. On the other hand, the MEC problem is known to be polynomial for a few special cases including bipartite graphs with edge weights $w(e) \in \{1, 2\}$ [2], chains [3] (in fact, this algorithm can be also applied for graphs of $\Delta = 2$), stars of chains and bounded degree trees [5].

Concerning the approximability of the MEC problem, a natural greedy 2-approximation algorithm for general graphs has been proposed by Kesselman and Kogan [4]. The ratio of this algorithm has been slightly improved to $2 - \frac{1}{\Delta}$ and $2 - \frac{2}{\Delta+2}$ in [6]. Especially for bipartite graphs of maximum degree $\Delta = 3$ an algorithm that attains the 7/6 inapproximability bound has been presented in [1]. For bipartite graphs have been also presented algorithms improving the best known $2 - \frac{2}{\Delta+2}$ approximation ratio for general graphs. In fact, algorithms presented in [3] and [5] achieve better ratios for bipartite graphs of $\Delta \leq 7$, and $\Delta \leq 12$, respectively. Moreover, two algorithms of approximation ratios $2 - \frac{2}{\Delta+1}$ and $\frac{2\Delta^3}{\Delta^3+\Delta^2+\Delta-1}$ have been presented in [6].

It is interesting that no algorithm of approximation ratio $2 - \delta$, for any small constant $\delta > 0$, is known for the MEC problem on bipartite graphs or even on trees. Recall that the MEC problem on bipartite graphs is 7/6 inapproximable and notice that neither the complexity of the MEC problem on trees is known. On the other hand, for the MVC problem this gap is closed for bipartite graphs and it is very narrow for trees. In fact, an algorithm which matches the inapproximability bound of 8/7 is known for bipartite graphs [2, 1, 7] while a PTAS is known for trees [7]. However, the complexity of the MVC problem on trees remains also unknown. In this note we decrease this gap for the MEC problem on trees by presenting a 3/2-approximation algorithm.

---

*Department of Informatics, Athens University of Economics and Business, Greece. `{gluc,milis}@aueb.gr`.
†LAMSADE, Université Paris-Dauphine, France. `paschos@lamsade.dauphine.fr`.

## 2  Definitions and Preliminaries

We consider the MEC problem on a weighted tree $T = (V, E)$. By $d(v)$ we denote the degree of vertex $v \in V$ and by $\Delta$ the maximum degree of $T$. By $S^* = \{M_1^*, M_2^*, \ldots, M_{s^*}^*\}$ we denote an optimal solution to the MEC problem of weight $OPT = w_1^* + w_2^* + \ldots + w_{s^*}^*$.

For each vertex $u \in V$, we denote by $E_u : e_1^u, e_2^u, \ldots, e_{d(u)}^u$ an ordering of its adjacent edges in non increasing weights, i.e. $w(e_1^u) \geq w(e_2^u) \geq \ldots \geq w(e_{d(u)}^u)$.

Furthermore, we define $y_i$, $1 \leq i \leq \Delta$, to be the weight of the heaviest edge between those ranked $i$ in each ordering $E_u, u \in V$, i.e. $y_i = \max_{u \in V}\{w(e_i^u)\}$. It is clear that $y_1 \geq y_2 \geq \ldots \geq y_\Delta$.

**Proposition 1** *For all $1 \leq i \leq \Delta$, it holds that $w_i^* \geq y_i$.*

**Proof:** Let $e = (u, v)$ be the heaviest edge with rank equal to $i$ i.e., $y_i = w(e)$. For at least one of the endpoints of $e$, assume w.l.o.g. for $u$, it holds that $e$ is ranked $i$ in $E_u$, that is $y_i = w(e_i^u)$. Therefore, there exist $i$ edges adjacent to vertex $v$ of weight at least $y_i$. These $i$ edges belong in $i$ different matchings in an optimal solution, since they share vertex $u$ as a common endpoint. Thus, the $i$-th matching in an optimal solution is of weight at least $y_i$. ∎

In [4], Kesselman and Kogan present the most interesting and general result we have for the MEC problem. This is a greedy 2-approximation algorithm for general graphs, to which we refer as Algorithm KK. A slightly better analysis of this algorithm presented in [6] leads to the following approximation ratio which also matches exactly the ratio of the tightness counterexample given in [4].

**Lemma 1** [6] *Algorithm KK achieves a tight approximation ratio of $2 - \frac{w_1^*}{OPT} < 2 - \frac{1}{\Delta}$.*

## 3  Approximation algorithm

In this section, we first present a $(1 + \frac{w_1^* - w_\Delta^*}{OPT})$-approximation algorithm for the MEC problem on trees.

Our algorithm roots the tree in an arbitrary vertex $r$ and constructs a solution as following: For each vertex $v \in V$, consider the edges to the children of $v$ in non increasing order and insert them into the first matching they fit.

**Algorithm 1**
```
1. Root the tree on an arbitrary vertex r;
2. For each vertex v in pre-order do
3.    Sort the children edges of v in non-increasing order,
      i.e.   w(e_1^v) ≥ w(e_2^v) ≥ ... ≥ w(e_{d(v)}^v);
4.    Using this order,
      insert each edge into the first matching that fits;
```

**Proposition 2** *Algorithm 1 constructs a solution of exactly $\Delta$ matchings. For the weight, $w_i$ of the $i$-th, $2 \leq i \leq \Delta$, matching it holds that $w_i \leq y_{i-1}$.*

**Proof:** For the first part of the lemma consider first the root vertex of the tree. It has at most $\Delta$ adjacent edges which the algorithm inserts into at most $\Delta$ different matchings. Consider, next, any other vertex $v$ and let $e$ be the edge between $v$ and its parent. This edge $e$ has been already inserted by the algorithm into a matching, say $M_k$. The rest, but $e$, adjacent to vertex $v$ edges are at most $\Delta - 1$ which the algorithm inserts into at most $\Delta - 1$ matchings different than $M_k$. Therefore, the algorithm will use exactly $\Delta$ matchings $M_1, M_2, \ldots, M_\Delta$.

We shall prove the second part of the lemma by induction on the vertices in the order they are processed by the algorithm.

For the root $r$, the algorithm sorts all adjacent edges to $r$ and inserts $e_1^r$ into matching $M_1$, $e_2^r$ into matching $M_2$, and so on. Thus, after the first iteration it holds that $w_i = w(e_i^r) \leq y_i \leq y_{i-1}$, $2 \leq i \leq \Delta$.

Assume that the statement of the lemma holds before the iteration processing the vertex $v \in V$, that is $w_i \leq y_{i-1}$, $2 \leq i \leq \Delta$.

Consider, now, the iteration in which the algorithm processes the vertex $v$. Let $e$ be the edge between $v$ and its parent, $j$ be the rank of the edge $e$ in $E_v$ and $M_k$ be the matching where the algorithm has already inserted edge $e$. Let us also denote by $w_i'$ the weight of the matching $M_i$, $2 \leq i \leq \Delta$, after processing the vertex $v$. We distinguish between three cases, and for each one we prove that $w_i' \leq y_{i-1}$, $2 \leq i \leq \Delta$.

(i) If $k = j$, then after this iteration each edge $e_i^v$ belongs to matching $M_i$, $1 \leq i \leq d(v)$. By the inductive hypothesis it follows that $w_i' = \max\{w_i, w(e_i^v)\}$, where $w_i \leq y_{i-1}$. Since $w(e_i^v) \leq y_i$, it holds that $w_i' \leq y_{i-1}$, $2 \leq i \leq \Delta$.

(ii) If $k > j$, then after this iteration: for $1 \leq i \leq j-1$ and $k+1 \leq i \leq d(v)$ each edge $e_i^v$ belongs to matching $M_i$; for $j+1 \leq i \leq k$ each edge $e_i^v$ belongs to matching $M_{i-1}$. For the former case we conclude as in Case (i). For the later case by the inductive hypothesis it follows that $w_i' = \max\{w_i, w(e_{i+1}^v)\}$ where $w_i \leq y_{i-1}$. Since $w(e_{i+1}^v) \leq y_{i+1} \leq y_{i-1}$, it holds that $w_i' \leq y_{i-1}$.

(iii) If $k < j$, then after this iteration: for $1 \leq i \leq k-1$ and $j+1 \leq i \leq d(v)$ each edge $e_i^v$ belongs to matching $M_i$; for $k \leq i \leq j-1$ each edge $e_i^v$ belongs to matching $M_{i+1}$. For the former case we conclude as in Case (i). For the later case by the inductive hypothesis it follows that $w_i' = \max\{w_i, w(e_{i-1}^v)\}$ where $w_i \leq y_{i-1}$. Since $w(e_{i-1}^v) \leq y_{i-1}$, it holds that $w_i' \leq y_{i-1}$. ∎

**Lemma 2** *Algorithm 1 achieves an approximation ratio equal to $1 + \frac{w_1^* - w_\Delta^*}{OPT}$ for the MEC problem on trees. This is an asymptotically tight 2 approximation ratio.*

**Proof:** For the weight of the first matching obtained by Algorithm 1 it holds that $w_1 \leq y_1 = w_1^*$, since both $y_1$ and $w_1^*$ are equal to the weight of heaviest edge of the tree. By Proposition 2 it holds that $w_i \leq y_{i-1}$, $2 \leq i \leq \Delta$ and by Proposition 1 it holds that $y_i \leq w_i^*$, $1 \leq i \leq \Delta$. Therefore, the weight of the solution obtained by Algorithm 1 is $W = \sum_{i=1}^{\Delta} w_i \leq y_1 + \sum_{i=2}^{\Delta} y_{i-1} = y_1 + \sum_{i=1}^{\Delta-1} y_i \leq w_1^* + \sum_{i=1}^{\Delta-1} w_i^* \leq w_1^* + OPT - w_\Delta^*$, that is $\frac{W}{OPT} \leq 1 + \frac{w_1^* - w_\Delta^*}{OPT} < 2$.

The counterexample in Figure 1(a) shows that this is an asymptotically tight 2 approximation ratio. The weight of an optimal solution to this instance is $C + 2\epsilon$ (Figure 1(b)) and the weight of the solution obtained by Algorithm 1 is $2C + \epsilon$ (Figure 1(b)). Thus, the approximation ratio for this instance becomes $\frac{2C+\epsilon}{C+2\epsilon}$. ∎

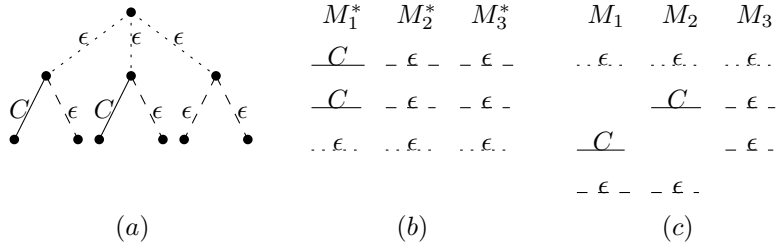

Figure 1: (a) A instance of the MEC problem where $C >> \epsilon$. (b) An optimal solution. (c) The solution obtained by Algorithm 1.

Next, we combine Algorithm KK [4] with Algorithm 1 i.e., we run both algorithms and we select the best of the two solutions found. For this new algorithm the next theorem holds.

**Theorem 1** *There is a tight $\frac{3}{2}$-approximation algorithm for the MEC problems on trees.*

**Proof:** Let $W$ be weight of the best of the two solutions found by

Algorithm KK and Algorithm 1. By Lemma 1 it holds that $\frac{W}{OPT} \leq 2 - \frac{w_1^*}{OPT}$ and by Lemma 2 that $\frac{W}{OPT} \leq 1 + \frac{w_1^* - w_\Delta^*}{OPT}$. As the first bound is increasing and the second one is decreasing with respect to $OPT$, it follows that the ratio $\frac{W}{OPT}$ is maximized when $2 - \frac{w_1^*}{OPT} = 1 + \frac{w_1^* - w_\Delta^*}{OPT}$, that is $OPT = 2 \cdot w_1^* - w_\Delta^*$. Therefore, $\frac{W}{OPT} \leq 2 - \frac{w_1^*}{OPT} = 2 - \frac{w_1^*}{2 \cdot w_1^* - w_\Delta^*} \leq 2 - \frac{w_1^*}{2 \cdot w_1^*} = \frac{3}{2}$.

For the tightness of this ratio consider the counterexample shown in Figure 2(a). The weight of an optimal solution to this instance is $2C + 2\epsilon$ (Figure 2(b)), the weight of the solution created by Algorithm 1 is $3C$ (Figure 2(c)) and the weight of the solution created by Algorithm KK (Figure 2(d)) is $3C - \epsilon$. Our algorithm selects the solution obtained by Algorithm KK of weight $3C - \epsilon$ and the approximation ratio for this instance becomes $\frac{3C - \epsilon}{2C + 2\epsilon}$. ∎

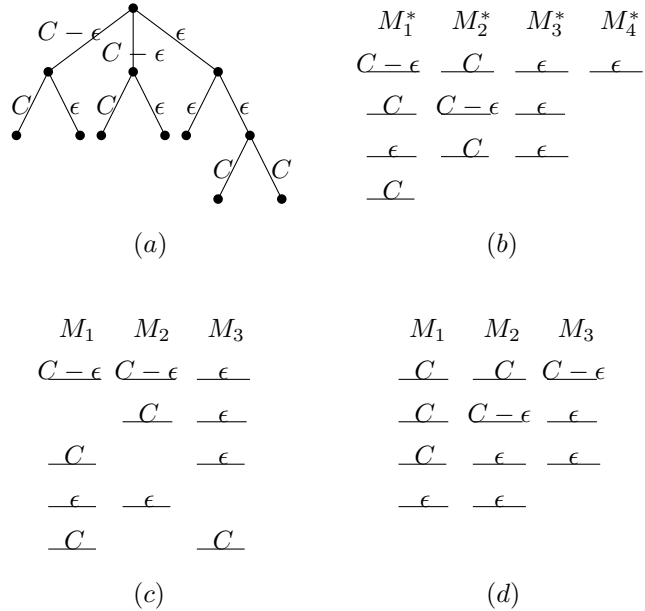

Figure 2: (a) A instance of the MEC problem where $C >> \epsilon$. (b) An optimal solution. (c) The solution obtained by Algorithm 1. (d) The solution obtained by Algorithm KK.